\documentclass[12pt]{article}
 \usepackage{amsmath,amsfonts,latexsym,amssymb,amsthm}
 \usepackage{times}

   \newcommand{\field}[1]{\mathbb{#1}}
   \newcommand{\rz}{\field{R}}
   \newcommand{\cz}{\field{C}}
   
   \newcommand{\nz}{\field{N}}

   \newcommand{\co}{C^\infty_0}
   \newcommand{\algnet}{\{{\cal A}(O)\}_{O \subset M}}
   \newcommand{\algO}{{\cal A}(O)}
   \newcommand{\Om}{\Omega}
   
   \newcommand{\rwz}{(\rz^n \backslash \{ 0\})}
   \newcommand{\vx}{\mathbf{x}}
   \newcommand{\vY}{\mathbf{Y}}

   \newcommand{\openset}{O}
   \newcommand{\Hilbert}{\mathcal{H}}
   \newcommand{\feld}{\mathcal{F}}
   \newcommand{\alg}{\mathcal{A}}
   \newcommand{\Di}{\mathcal{D}'{}}
   \newcommand{\DiE}{\mathcal{E}'{}}
   \newcommand{\WF}{\mathrm{WF}}
   \newcommand{\I}{\mathrm{i}}
   \newcommand{\U}{\mathcal{U}}
   \newcommand{\nts}[1]{\tmspace{-}{#1\thinmuskip}{#1\txtmu}}
   \newcommand{\dint}{\int \nts{2.5} \int}
   \newcommand{\msc}{$\mu \mathrm{SC}$}
   \newcommand{\amsc}{$a\mu \mathrm{SC}$}
   \newcommand{\supp}{\mathrm{supp}}
   
 \newtheorem{theorem}{Theorem}[section]
 \newtheorem{definition}[theorem]{Definition}
 
 \newtheorem{lem}[theorem]{Lemma}
 \newtheorem{cor}[theorem]{Corollary}
 \newtheorem{pro}[theorem]{Proposition}
 
 \newtheorem{rem}[theorem]{Remark}

\begin{document}

\title{Microlocal analysis of quantum fields on curved
                  spacetimes: Analytic wavefront sets and
                  Reeh-Schlieder theorems}

\author{Alexander Strohmaier$^{a}$, Rainer Verch$^{b}$ \\
and Manfred Wollenberg$^{c}$ }

\date{\small
$^a$ Institut f\"ur Theoretische Physik, Universit\"at Leipzig \\
Augustusplatz 10--11, D--04109 Leipzig, Germany\\
$^b$ Institut f\"ur Theoretische Physik, Universit\"at G\"ottingen\\
Bunsenstr. 9, D--37073 G\"ottingen, Germany\\
$^c$ Institut f\"ur Mathematik, Universit\"at Leipzig \\
Augustusplatz 10--11, D--04109 Leipzig, Germany
}

\maketitle
\noindent
\begin{abstract}
  \noindent
 We show in this article that the Reeh-Schlieder property holds for states of
 quantum fields on real analytic curved spacetimes if they satisfy
 an analytic microlocal spectrum condition.
 This result holds in the setting of general quantum field theory,
 i.e.\ without assuming the quantum field to obey a specific equation
 of motion. Moreover,
 quasifree states of the
 Klein-Gordon field are further investigated in the present work
  and the (analytic) microlocal
 spectrum condition is shown to be equivalent to simpler conditions.
 We also prove that any quasifree ground- or KMS-state
 of the Klein-Gordon field on a stationary real analytic spacetime
 fulfills the analytic microlocal spectrum condition.
\end{abstract}

{\small \bf Mathematics Subject Classification (2000):} 81T05, 81T20, 35A18, 35A20
\\

\maketitle

\section{Introduction}
One of the remarkable features of quantum field theory is the ubiquity
of fluctuations and, connected with that, the generic appearance of
long-range correlations. What is even more remarkable is the fact
that, using suitable selective operations and applying them, say, in
an arbitrary spacetime region to the vacuum, one may produce in this
way any given state in any other causally separated spacetime region
up to arbitrary precision. This is known as the Reeh-Schlieder theorem
\cite{Reeh:1961re}. Let us recall its statement in the setting of the
operator-algebraic approach to quantum field theory. Suppose that we
are given a spacetime manifold $(M,g)$ 
\footnote{i.e.\ $M$ is a smooth,
$n$-dimensional manifold, and $g$ is a Lorentzian metric on $M$}
and a family (``local net'') $\algnet$ of von Neumann algebras, all acting on a
Hilbert-space $\Hilbert$; the family is indexed by the open,
relatively compact subsets of $M$, and is subject to the conditions of
isotony and locality:
$$ O_1 \subset O \Rightarrow \alg (O_1) \subset \algO \quad {\rm and}
\quad O_1 \subset O^{\perp} \Rightarrow \alg (O_1) \subset \algO'\,.$$
Here, $O^{\perp}$ denotes the causal complement of $O$, i.e.\ the set
of all points in $M$ which cannot be connected to $O$ by any causal
curve, and $\algO'$ denotes the commutant algebra of $\algO$ in
$B(\Hilbert)$. These conditions are the minimal assumptions in order
that $\algnet$ may be viewed as a net of local observable algebras of
a (relativistic) physical system situated in $M$, see \cite{Haag:1992hx} for
discussion. One now says that a unit vector $\Om \in \Hilbert$
satisfies the Reeh-Schlieder property with respect to the region
$O\subset M$ if $\Om$ is cyclic for the algebra $\algO$ of observables
localized in $O$, that is, the set of vectors $\algO \Om = \{A\Om : A
\in \algO\}$ is dense in $\Hilbert$. Moreover, one says that $\Om$ has
the Reeh-Schlieder property if $\Om$ is cyclic for $\algO$ for each $O
\subset M$ which is open, non-void, and relatively compact. By the locality
assumption, this then implies that $\Om$ is also separating for all
local algebras $\algO$ (as long as $O^{\perp}$ contains a non-void
open set) and this means that $A\Om = 0 \Rightarrow A = 0$ for all $A
\in \algO$. (A vector $\Om \in \Hilbert$ which is cyclic and separating
for all local algebras $\algO$ is sometimes also called a standard
vector for the family $\algnet$.) 
\par
The generic occurrence of the Reeh-Schlieder property for large sets of
physical states in quantum field theory --- as so far known in quantum
field theory on manifolds possessing suitable groups of isometries
\cite{Reeh:1961re,SW,BW,D'Antoni:1989pv,Jakel:1999ji,Verch:1993pn,Strohmaier:2000ye,BaerS:2001} 
--- is a 
mathematically precise way of expressing that long-range correlations
are a fundamental feature of quantum field theory. Furthermore, the
Reeh-Schlieder property plays a very important role in analyzing the
mathematical structure of quantum field theory. For instance, it is
being used at some stages in the development of charge superselection
theory (see \cite{BW,Roberts:1989ps} and references cited there). Another
very important aspect of the Reeh-Schlieder property is that one may
naturally associate with each 
von Neumann algebra together with a cyclic and separating vector 
the so-called Tomita-Takesaki modular objects \cite{Ta70}. In the
seminal work of Bisognano and Wichmann
\cite{Bisognano:1975ih,Bisognano:1976za}
it has been shown
that the Tomita-Takesaki modular objects associated with the
vacuum-vector and the von Neumann algebra $\alg (W)$ of a
``wedge-region'' (which is actually infinitely extended) in a
Wightman-type quantum field theory on Minkowski spacetime have a
specific geometric significance. This insight has initiated considerable
progress in the mathematical development of general quantum field
theory on which the recent review by Borchers \cite{Borchers:2000pv} reports
exhaustively; therefore we refer the reader to that reference for
further discussion. We limit ourselves to mentioning that quite
promising generalized forms of such a ``geometric modular action''
that are applicable to quantum field theories on curved spacetimes
have been suggested and investigated more recently 
\cite{Buchholz:1998pv,Buchholz:2000ws}.
The Reeh-Schlieder property is also responsible for (maximal)
violations of Bell's inequalities in quantum field theory 
\cite{SumWer}, and  
more recently, Reeh-Schlieder properties have been found to imply
various forms of long-range entanglement of states
in relativistic quantum field theory 
\cite{Halvorson:1999pz,Jakel:2001az,VerWer}; see
also \cite{Summers:1990tp,Beckman:2001qs} for related discussions. A possible significance of
Reeh-Schlieder properties for questions related to cosmology has been
proposed in \cite{Wald:1992wf}.  
\par
As indicated above, Reeh-Schlieder properties have, either in
the model-inde\-pendent approach or for concrete quantum field models,
so far only been established under the assumption that the spacetime
in which the quantum system is situated possesses a sufficient amount
of spacetime symmetries. This constitutes a considerable limitation,
and the question is if Reeh-Schlieder like properties can also be
established for quantum field theories on spacetimes not admitting any
isometries. This is feasible since the main mathematical argument
leading to the Reeh-Schlieder theorem in the case where there are
sufficiently many spacetime symmetries is an analytical argument of the type
of the edge-of-the-wedge-theorem \cite{SW} or Schwartz' reflection
principle in order to derive a certain global property of a quantum state
from local information, and these arguments don't use
spacetime-symmetries (in particular, timelike isometry groups)
directly. On the other hand, commonly the analytic properties of
correlation functions in quantum field theory are consequences of the
relativistic spectrum condition whose formulation requires a form of
time-translation symmetry. Time-translation symmetry is also required
in order to formulate conditions of thermal equilibrium in
relativistic quantum field theory from which Reeh-Schlieder properties
may also be deduced \cite{Jakel:1999ji}. 
\par
At any rate, certain analytical properties of correlation functions
are prerequisite in order to establish Reeh-Schlieder theorems in
quantum field theory and the question arises how to generalize the
analyticity properties known to hold e.g.\ for ground states or
thermal equilibrium states with respect to a time-symmetry group to
more general situations. A way how to proceed in more general
situations may be to follow, and to refine, the approach pioneered by
Radzikowski \cite{Radzikowski:1992eg,Radzikowski:1996pa}
who proved that, for a free scalar quantum
field on generic curved spacetimes, the condition that a quasifree
state be a Hadamard state can equivalently be expressed as a condition
on the wavefront set of the two-point function of that state (see also
\cite{Kratzert:2000bn,Hollands:1999fc,Sahlmann:2000ij} for related work). Since it can
convincingly be argued that Hadamard states are most likely candidates
for physical states of quantum fields obeying linear wave-equations
\cite{Wald:1977up}, it appears natural to sharpen the condition on the
two-point function of a physical state by demanding that it applies to
the analytic wavefront set and not only, as in most previous
considerations, to the $C^{\infty}$-wavefront set. In order that this
makes sense independently of particular coordinates, the underlying
space-time manifold ought to be real analytic.
\par
 Thus, we will propose a
stricter form of the wavefront set spectrum condition formulated in
\cite{Radzikowski:1992eg}, or of the microlocal
spectrum condition \cite{Brunetti:1996rf}, for
the $n$-point functions of a generic scalar quantum field on a
real analytic spacetime in terms of their analytic
wavefront sets, and we will show that such states possess the
Reeh-Schlieder property. In doing so, we will present a transcription
of the (analytic) microlocal spectrum condition for a two-point
function as a condition on the wavefront set of a certain Hilbert-space valued
distribution. It turns out that working with wavefront sets of
Hilbert-space valued distributions has several advantages. One of
those is a that in terms of Hilbert-space valued distributions the
(analytic) microlocal spectrum condition (for two-point functions of
free fields)
assumes a very simple and elegant form which is, in fact, most
reminiscent of the usual spectrum condition in quantum field theory in
Minkowski spacetime. This observation appeared motivation enough to
discuss several aspects of microlocal analysis of Hilbert-space valued
distributions more systematically, and that discussion thus forms the
first part of the present article in Sec.\ 2.

In Sec.\ 3 we summarize some basics on the description of general
scalar quantum fields on curved spacetimes, together with the example
of the free scalar Klein-Gordon field. 

We recapitulate the definition of the microlocal spectrum condition
(``$\mu SC$''), referring to the $C^\infty$-wavefront sets of
$n$-point functions of quantum fields on manifolds, according to
\cite{Brunetti:1996rf} in Sec.\ 4. In the same section, we introduce our analytic
microlocal spectrum condition (``$a\mu SC$'') which will be defined
similarly to $\mu SC$ but using now analytic wavefront sets of
$n$-point functions of quantum fields on real analytic manifolds. (We
should note that Hollands and Wald \cite{HollandsW:2001} have recently also
introduced a similar concept of analytic microlocal spectrum condition which
refers to a whole class of states, and is used in a different context).

In Sec.\ 5 we will present our main result, which is a Reeh-Schlieder
theorem for quantum field states fulfilling the $a\mu SC$. Here, we
draw on results of Sec.\ 2, and also on a microlocal version of the
edge-of-the-wedge theorem, which appears as Thm.\ 8.5.6' in
\cite{HormBook:1990}.

Finally, in Sec.\ 6 we present the characterization of $a\mu SC$ and
of $\mu SC$ for 
two-point functions of a free scalar field on a manifold in terms of a
simple conic form of the wavefront set of certain Hilbert-space valued
distributions. Moreover, we show that ground- and KMS-states of the
free scalar field on a real analytic stationary spacetime obey the
$a\mu SC$.    

\section{Microlocal analysis for Hilbert space valued distributions}

Assume we are given a Hilbert space $\Hilbert$ and a smooth manifold $M$
which is second countable and Hausdorff.
The space $\Di(M,\Hilbert)$ of $\Hilbert$ valued distributions is defined to be
the set of all weakly continuous linear maps
$$
\co(M) \to \Hilbert.
$$
Note that due to the nuclearity of $\co(M)$ these maps are automatically
strongly continuous.
If $X \subset \rz^n$ is an open subset a linear map
$\psi: \co(X) \to \Hilbert$ is in $\Di(X,\Hilbert)$ if and only if for
each compact subset $K \subset X$ there are constants $C>0$ and
$\alpha \in \nz_0$ such that
\begin{equation}
 \Vert \psi(f) \Vert < C \sum_{\vert k \vert \leq \alpha}
 \sup_{x} \vert (\partial^k f)(x) \vert,
\end{equation}
for all $f \in \co(K)$.
If $\alpha$ can be chosen independently of $K$, then we say that
$\psi$ is of order $\alpha$ and write $\psi \in \Di^\alpha(X,\Hilbert)$.
The set of $\psi \in \Di(X,\Hilbert)$ with compact support in $X$ will be denoted
by $\DiE(X,\Hilbert)$. If $\psi$ has compact support then one can easily
extend $\psi$ to a linear map $C^\infty(X) \to \Hilbert$ and we have
for all compact sets $K$ which contain $\textrm{supp}(\psi)$
\begin{equation} \label{seminorm}
 \Vert \psi(f) \Vert < C \sum_{\vert k \vert \leq \alpha}
 \sup_{x \in K} \vert (\partial^k f)(x) \vert, \quad \forall f \in C^\infty(X),
\end{equation}
for some $C>0, \alpha \in \nz_0$.
Conversely, if there exists a compact set $K$ such that for a given linear map
$\psi: C^\infty(X) \to \Hilbert$ the inequality (\ref{seminorm}) holds for
some $C$ and $\alpha$, then $\psi$ is a distribution with support in $K$.
Therefore, $\DiE(X,\Hilbert)$ can be identified with the set of strongly continuous maps
$C^\infty(X) \to \Hilbert$.
If a subset $L$ of $\DiE(X,\Hilbert)$ is bounded then (\ref{seminorm})
holds for all $\psi \in L$ and $f \in C^\infty(X)$ with constants $C$ and
$\alpha$ independent of $\psi$. We define
$\DiE^\alpha(X,\Hilbert):=\DiE(X,\Hilbert) \cap \Di^\alpha(X,\Hilbert)$
and obviously $\DiE(X,\Hilbert)=\bigcup_{\alpha} \DiE^\alpha(X,\Hilbert)$.

If $\psi \in \DiE(\rz^n,\Hilbert)$ one may define the
Fourier transform $\hat \psi$ in the same way as this is done for ordinary
distributions. Namely, the Fourier transform $\hat \psi$ is the $\Hilbert$ valued function on $\rz^n$
given by $\hat \psi (k) := \psi(e^{-\I k \cdot})$.

\begin{definition}
 Let $X$ be open in $\rz^n$ and let $\psi$ be in $\Di(X,\Hilbert)$. Then a point
 $(x,k) \in X \times \rwz$ is called regular directed
 for $\psi$ if the following holds:
 There exists a function $f \in \co(X)$ with $f(x)=1$ and an open conic
 neighbourhood $\Gamma$ of $k$ such that for each $N \in \nz$
 there exists a constant $C_N$ with
 \begin{equation} \label{inequ}
  \sup_{\lambda \in \Gamma} (1 + \vert \lambda \vert)^N \Vert \widehat{f \psi}(\lambda) \Vert < C_N.
 \end{equation}
 The set of regular directed points is open. Its complement in
 $X \times \rwz$ is called the
 wavefront set $\textrm{WF}(\psi)$ of $\psi$.
\end{definition}

The following proposition shows that microlocal analysis of Hilbert
space valued distributions is analogous to the case of ordinary
distributions.

\begin{pro}\label{m1}
  Let $X$ be open in $\rz^n$ and let $\psi \in \Di(X,\Hilbert)$ be a Hilbert
  space valued distribution.
  \begin{enumerate}
   \item[1)] If (\ref{inequ}) holds for $f \in \co(X)$ then it also holds
         with $f$ replaced by $g f$ for any $g \in C^\infty(X)$.
   \item[2)] If $\psi$ has compact support then $\hat \psi$ is polynomially
         bounded in the norm, i.e. there is a constant $C$
         and an integer $M$, such that
         \begin{equation}\label{polybound1}
          \Vert \hat \psi (k) \Vert < C (1+\vert k \vert)^M.
         \end{equation}
         If moreover a subset $L \in \DiE(X,\Hilbert)$ is bounded then
         (\ref{polybound1}) holds for all $\psi \in L$ with $C$ and $M$
         independent of $\psi.$
   \item[3)] If $\WF(\psi)$ is empty then $\psi$ is smooth in the norm.
   \item[4)] If we define the distribution $w \in \Di(X \times X)$
         by $w(f,g)=\langle \psi(\overline f), \psi(g) \rangle$, then
         \begin{equation}
            (x,k) \in \textrm{WF}(\psi) \Leftrightarrow  ((x,-k),(x,k)) \in
            \textrm{WF}(w),
         \end{equation}
         and moreover, if $(x,k) \notin \textrm{WF}(\psi)$, then
         \begin{equation}
          ((x,-k),(x_1,k_1)) \notin \textrm{WF}(w) \textrm{ and }
          ((x_1,k_1),(x,k)) \notin \textrm{WF}(w),
         \end{equation}
         for arbitrary
         $(x_1,k_1) \in X \times \rz^n$.
    \item[5)] Under change of coordinates $\WF(\psi)$ transforms as a subset
         of the cotangent bundle. Hence, $\WF(\phi)$ may be defined for
         distributions $\phi \in \Di(M,\Hilbert)$, for a smooth manifold
         $M$ and $\WF(\phi) \subset T^*M \backslash 0$. Here
         $T^*M \backslash 0$ is the cotangent bundle with zero section removed.
         
  \end{enumerate}
\end{pro}

\begin{proof}
    The first statement is proved in the same manner as for ordinary
    distributions (cf. \cite{HormBook:1990}, Lemma 8.1.1).
    The inequality (\ref{seminorm}) immediately gives the second statement
     if $f(x)=e^{-\I k x}$.
     To see that 3) holds we first note that
     $\WF(\psi)=\emptyset$ implies that for each point
     $x \in X$ there is a positive function with $f(x)=1$ such that 
     $\Vert \widehat{ f \psi} \Vert \in \mathcal{S}(\rz^n)$. The same arguments as in
     the complex valued case show that $f \psi$ can be represented by the inverse
     Fourier transform of $\widehat{ f \psi}$, which is a smooth function, i.e.
     all derivatives in the norm sense exist and are given by
     $D^\alpha (f \psi) (x) = 2 \pi \widehat{((-\I k)^\alpha \widehat{f \psi}(k)}(-x)$.
     5) is a simple consequence of 4) and it remains to
     show 4). Assume first that $(x,-k,x,k)$ is regular directed for $w$. Then there is
     a function $f_1 \in \co(X \times X)$ such that 
     $\widehat{f_1 w}$ decays rapidly in a conic neighbourhood of $(x,-k,x,k)$.
     Because of 1) we may choose $f_1$ to be of the form $f \otimes f$, where
     $f \in \co(X)$ is a positive function. Since
     \begin{equation}
      \Vert \widehat{f \psi}(k)  \Vert^2 = \widehat{\left((f \otimes f) w\right)}(-k,k),
     \end{equation}
     $(x,k)$ is regular directed for $\psi$. Suppose conversely we knew that
     $(x,k)$ is regular directed for $\psi$. 
     By the Cauchy inequality we have
     \begin{equation}\label{inequ2}
       \vert (\widehat{(f \otimes g)w})(k_1,k_2) \vert \leq
      \Vert \psi(\overline f(\cdot) e^{\I k_1 \cdot}) \Vert \cdot \Vert
      \psi(g(\cdot) e^{-\I k_2 \cdot}) \Vert.
     \end{equation}
     By assumption there is a $g \in \co(X)$ with $g(x)=1$ such that in
     a conic neighbourhood of $k$ the second factor is rapidly decreasing.
     Since the other is polynomially bounded for any $f \in \co(X)$,
     $(x_1,k_1,x,k)$ is a regular directed point for $w$ for any $(x_1,k_1)$.
     In the same way one shows that $(x,-k,x_1,k_1)$ is regular
     directed. This concludes the proof.
\end{proof}

The fourth statement in the above proposition allows one to
take over many results known for ordinary distributions.

\begin{pro}\label{diffwf}
 If $P: C^\infty(M) \to C^\infty(M)$ is a differential operator and
 $\psi \in \Di(M,\Hilbert)$ such that $\psi \circ P^* = 0$,
 where $P^*$ is the formal adjoint of $P$, then
 $$
  \textrm{WF}(\psi) \subset \textrm{char}(P).
 $$
 Here $\textrm{char}(P)$ is the characteristic set of $P$, i.e.
 the set of points $(x,k)$  in $T^*M \backslash 0$ on which the
 principal symbol $\sigma_P$ of $P$ vanishes.
\end{pro}

\begin{proof}
 We define $w$ as in Proposition \ref{m1}. 
 Note that $((x,-k),(x,k))$ is in the characteristic set of the operator
 $L:=\overline P \otimes P$ if and only if $(x,k)$ is in the characteristic
 set of $P$. Moreover, $L w = 0$.
 The result follows now from the fourth statement in Proposition \ref{m1}.
\end{proof}

One may also define the analytic wavefront set of a Hilbert space valued
distribution. We follow the definition in \cite{HormBook:1990} (Def.
8.4.3).

\begin{definition}
 Let $X$ be an open subset of $\rz^n$ and $\psi \in \Di(X,\Hilbert)$.
 We denote by $\WF_A(\psi)$ the complement in
 $X \times (\rz^n \backslash \{ 0 \})$ of the set of points $(x_0,k_0)$
 such that there is a neighbourhood $U \subset X$ of $x_0$ , a conic
 neighbourhood $\Gamma$ of $k_0$ and a bounded sequence $\psi_N$
 of distributions with compact support which is equal to $\psi$
 in $U$, such that there exists a constant $C$ with
 \begin{equation}\label{dawf}
  \Vert \vert k \vert^N \hat \psi_N (k) \Vert \leq C (C (N+1) )^N,
 \end{equation}
 for all $k \in \Gamma$.
\end{definition}

The bounded sequence $\psi_N$ can always be chosen to be the product
$f_N \psi$, where $f_N$ is a sequence of smooth functions. One has
\begin{lem}
 Let $\psi \in \Di(X,\Hilbert)$, $K$ a compact subset of $X$, and let
 $F$ be a closed cone in $\rz^n$ such that
 $\WF_A(\psi) \cap (K \times F) = \emptyset$.
 If $f_N \in \co(K)$ and for all $\alpha$
 \begin{equation} \label{regfunc}
  \vert D^{\alpha +\beta} f_N
  \vert \leq C_\alpha (C_\alpha (N+1))^{\vert \beta \vert} ,
  \quad \vert \beta \vert \leq N = 1,2, \ldots ,
 \end{equation}
 then $f_N \psi$ is a bounded sequence and we have
 \begin{equation}
   \Vert \vert k \vert^N \hat \psi_N (k) \Vert \leq C (C (N+1) )^N,
 \end{equation}
 for all $k \in F$.
 Moreover, if $x$ is a point in the interior of $K$, there always exists
 a neighbourhood $\U$ of $x$ and a sequence of functions $f_N$ such that
 (\ref{regfunc}) is satisfied and $f_N=1$ on $\U$.
\end{lem}

\begin{proof}
 The proof of this statement is the same as for ordinary distributions (see
 \cite{HormBook:1990}, Lemma 8.4.4).
\end{proof}

\begin{pro}\label{m2}
 Let $X$ be an open set in $\rz^n$ and let
 $\psi \in \Di(X,\Hilbert)$ be a Hilbert space valued distribution.
  \begin{enumerate}
   \item[1)] If $\WF_A(\psi)$ is empty then $\psi$ is strongly real analytic. 
   \item[2)] If we define the distribution $w \in \Di(X \times X)$
         by $w(f,g)=\langle \psi(\overline f), \psi(g) \rangle$, then
         \begin{equation}
            (x,k) \in \textrm{WF}_A(\psi) \Leftrightarrow  ((x,-k),(x,k)) \in \textrm{WF}_A(w),
         \end{equation}
         and moreover, if $(x,k) \notin \textrm{WF}_A(\psi)$, then
         \begin{equation}
          ((x,-k),(x_1,k_1)) \notin \textrm{WF}_A(w) \textrm{ and }
          ((x_1,k_1),(x,k)) \notin \textrm{WF}_A(w),
         \end{equation}
         for arbitrary
         $(x_1,k_1) \in X \times \rz^n$.
    \item[3)] Under analytic change of coordinates $\WF_A(\psi)$ transforms as a subset
         of the cotangent bundle. Hence, $\WF_A(\phi)$ may be defined for
         distributions $\phi \in \Di(M,\Hilbert)$, for a smooth real analytic
         manifold $M$ and $\WF_A(\phi) \subset T^*M \backslash 0$.
  \end{enumerate}
\end{pro}

\begin{proof}
 We start with 2). Assume that $(x,-k,x,k) \notin \WF_A(w)$.
 We choose a sequence of functions $f_N$ in $\co(X)$ which satisfies
 the inequality (\ref{regfunc}) and which is equal to $1$ in a neighbourhood
 of $x$. Then the sequence $g_N:=\overline f_N \otimes f_N \in \co(X \times X)$
 also satisfies an inequality of the form (\ref{regfunc}), and hence
 \begin{equation}
   \vert (-k_1,k_1) \vert^N \vert \widehat{ (g_N w)} (-k_1,k_1) \vert \leq C (C (N+1) )^N,
 \end{equation}
 for some constant $C > 0$ and all $k_1$ in a conic neighbourhood
 of $k$.
 We have
 \begin{equation}
   \Vert \widehat{f_N \psi}(k_1)\Vert^2 = \widehat{(g_N w)}(-k_1,k_1),
 \end{equation}
 and a quick estimate shows that for all $k_1$ in a conic neighbourhood of
 $k$ we have
 \begin{equation}
   \Vert \vert k_1 \vert^N \widehat{ (f_N \psi)} (k_1) \Vert \leq \tilde C
   (\tilde C (N+1) )^N,
 \end{equation}
 for some constant $\tilde C$. Therefore, $(x,k) \notin \WF_A(\psi)$.
 Suppose conversely that $(x_1,k_1) \notin \WF_A(\psi)$. Hence, there is
 a sequence $\psi_N$ bounded in $\DiE(X,\Hilbert)$ and equal
 to $\psi$ in a neighbourhood of $x_1$ such that the inequality (\ref{dawf}) holds
 in a conic neighbourhood of $k_1$.
 Choose another function $g$ which is equal to $1$ in a neighbourhood
 of a point $x_2$. Then the distribution $w_N$ defined by
 $w_N(h_1,h_2):=\langle \psi(g \cdot \overline h_1),\psi_N(h_2) \rangle$
 is bounded in $\DiE(X \times X)$. Moreover, an application
 of the Cauchy inequality shows that
 \begin{equation}
   \vert \hat w_N(k_2,k) \vert \leq \Vert \psi(g e^{\I k_2 \cdot})  \Vert
   \cdot \Vert \psi_N(e^{-\I k \cdot}) \Vert.
 \end{equation}
 The first factor is bounded by $C_M (1+\vert k_2 \vert)^M$ for some $M \in \nz$ 
 and a simple estimate shows that for all $k_2$ there is a conic neighbourhood
 $\Gamma$ of $(k_2,k_1)$ and a $C > 0$  with
 \begin{equation}
  \vert (k_2',k)\vert^N \cdot \vert \hat w_{N+M}(k_2',k) \vert
  \leq C \left ( C (N+1) \right )^N,
 \end{equation}
 for all $(k_2',k) \in \Gamma$.
 Since $w_N$ is equal to $w$ in a neighbourhood of $(x_2,x_1)$, we get
 $$((x_2,k_2),(x_1,k_1)) \notin \textrm{WF}_A(w).$$ In the same way one shows that
 $$((x_1,-k_1),(x_2,k_2)) \notin \textrm{WF}_A(w).$$
 Statement 3) is an immediate consequence of 2) since the analytic wavefront set
 of an ordinary distribution transforms as a subset of the cotangent bundle.
 The first statement can be shown in the same way as for ordinary distributions
 (see \cite{HormBook:1990}, Theorem 8.4.5). This concludes the proof.
\end{proof}

\begin{pro} \label{adiffwf}
 If $P: C^\infty(M) \to C^\infty(M)$
 is a differential operator with real analytic
 coefficients on a real analytic manifold $M$ and
 $\psi \in \Di(M,\Hilbert)$ such that $\psi \circ P^* = 0$,
 where $P^*$ is the formal adjoint of $P$, then
 $$
  \textrm{WF}_A(\psi) \subset \textrm{char}(P),
 $$
 where $\textrm{char}(P)$ is the characteristic set of $P$.
\end{pro}

\begin{proof}
  Analogous to the proof of Proposition \ref{diffwf}.
\end{proof}

\begin{theorem}\label{boundval}
 Let $X \subset \rz^n$ be an open subset and assume
 $\psi \in \Di(X,\Hilbert)$. Assume that there is a smooth
 $\Hilbert$-valued function $G: I \times X \to \Hilbert$, with $I=(0,\epsilon)$,
 such that
 \begin{itemize}
   \item $\lim_{t \to 0} G(t,\cdot) = \psi$ in the sense of distributions,
   \item $(\partial_t G)(t,x_1,\ldots,x_n)=\I (\partial_{x_1} G)(t,x_1,\ldots,x_n)$.
 \end{itemize}
 Then $\textrm{WF}_A(\psi) \subset \{ (x_1,\ldots,x_n) \times (y_1,\ldots,y_n)
 \in X \times \rz^n \backslash \{0\}; y_1 \geq 0 \}$.
\end{theorem}

\begin{proof}
 The proof is a variation of the proof of Theorem 8.4.8 in \cite{HormBook:1990}.
 Since the statement is local we can assume without loss of generality
 that $X=X_1 \times \ldots \times X_n$ and that $\psi$.
 First note that for each given $g \in \co(X_2 \times \ldots \times X_n)$ the function
 \begin{equation}
  H(x+\I y):=\int_{X_2 \times \ldots \times X_n} G(y,x,y_2,\ldots,y_{n})
  g(y_2,\ldots,y_n) dy_2 \cdots dy_n
 \end{equation}
 is defined on the strip $X_1 + \I I$ and is holomorphic.
 Moreover, it has a distributional boundary value $\psi(\cdot \otimes g)$.
 We will slightly vary the proof of theorem 3.1.14 in \cite{HormBook:1990}
 to show that the following bound holds:
 \begin{equation}\label{polybound}
  \Vert H(x+\I y) \Vert \leq C' y^{-m-1}
 \end{equation}
 if $(x,y) \in \tilde X_1 \times I/2$ and $\textrm{clo}(\tilde X_1) \subset
 X_1$ for some $m>0$.
 Let $f \in \co(X_1 \times I)$ be a function with support in $K \times I$,
 where $K$ is a compact subset of $X_1$, such that $f$ is  
 equal to one in a neighbourhood of $\overline Z$, where  $Z:=\tilde X_1
 \times I/2$.  Cauchy's integral formula applied to $f H$ in the set
 $\Im z > \Im (\zeta/2)$ shows that if $\zeta=\xi + \I \eta \in Z$
 \begin{eqnarray}\label{abc} \nonumber
   H(\zeta)& =-&\pi^{-1} \dint_{y > \eta/2} H(x+\I y) \partial f(x,y)/
   {\partial 
   \overline z} (z-\zeta)^{-1} dx{} dy\\
 & +& (2 \pi \I)^{-1} \int f(x,\eta/2)(x-\xi-\I\eta/2)^{-1} H(x+\I
 \eta/2)\; dx. 
 \end{eqnarray}
 An application of the uniform boundedness principle (Banach-Steinhaus
 theorem) shows that
 $\Vert \int_{X_1} H(x +\I y) h(x) dx \Vert \leq C \sum_{\alpha \leq m} \sup \vert \partial^\alpha h
 \vert$ for all $h \in \co(K)$ with constants $C$ and $m$ independent of
 $y$ (c.f. \cite{HormBook:1990}, Theorem 2.1.8).
 Therefore, the last integral in (\eqref{abc}) can be estimated in the norm by
 \begin{gather}
  C_1 \sum_{\alpha \leq m}
  \sup \vert \partial_x^\alpha(x,\eta/2)(x-\xi-\I\eta/2)^{-1}\vert
  \leq C_2 \vert \eta \vert^{-m-1}.
 \end{gather}
 The first integral in \eqref{abc} is bounded and this proves the
 inequality (\ref{polybound}). 
 
 Since the statement of the theorem is local we can always replace
 $X_1$ by $\tilde X_1$ and $I$ by $I/2$ 
 and we can therefore assume without loss of generality that the bound
 (\ref{polybound}) holds in $X_1 \times 
 I$.
 From Stokes integral formula one gets for $y,Y \in \rz_+$ and $N \in \nz$ such that
 $y+Y < \epsilon/2$ the following formula (compare 3.1.19 in
 \cite{HormBook:1990}) for any $\tilde \phi \in \co(X)$ with $\tilde
 \phi=\phi \otimes g$: 
 \begin{eqnarray} \label{intform}
   \lefteqn{\int_X \tilde \phi(\vx) G(y,\vx) d\vx = \int_{X_1}
   \Theta(x,Y) H(x+\I y+\I Y) dx}  \\ \nonumber
   & + &(N+1) \int_{X_1} dx \int_{0<t<1} dt\; H(x+\I tY+\I y)
   (\partial^N \phi)(x) \frac{(\I Y)^N}{N!} t^N,
 \end{eqnarray}
 where
 \begin{equation}
  \Theta(x,y):= \sum_{j=0}^N \partial^j \phi (x) (\I y)^j / j!\;.
 \end{equation}
 Because of the bound (\ref{polybound}) the integrand under the double
 integral in (\ref{intform}) is uniformly bounded by an integrable
 function if $N > (m+1)$ and the first term even converges uniformly
 as $y \to 0$. 
 Therefore, we have for $N > (m+1)$
 \begin{eqnarray}\label{intform2}
 \lefteqn{\psi(\tilde \phi) = \int_{X_1} \Theta(x,Y) H(x+\I Y) dx}
  \\ \nonumber
   & + & (N+1) \int_{X_1} dx \int_{0<t<1} dt\; H(x+\I tY) (\partial^N \phi)(x)
   \frac{(\I Y)^N}{N!} t^N .
 \end{eqnarray}
 Now let $\vx'=(x_1',\ldots,x_n')$ be a point in $X$ and let
 $\phi_\nu$ be a sequence of functions 
 on $X_1$ which are all equal to $1$ in a common neighbourhood of $x_1'$ such
 that
 \begin{equation}
  \vert \partial^\alpha \phi_\nu \vert \leq (C_1(\nu+1))^\alpha, \quad \alpha
  \leq \nu + 1.
 \end{equation}
 Assume that $g$ is equal to $1$ in a neighbourhood of $(x_2',\ldots,x_n')$.
 With
 \begin{equation}
   \Theta_\nu(x,y):= \sum_{j=0}^\nu \partial^j \phi_\nu (x) (\I y)^j / j!
 \end{equation}
 we get from (\ref{intform2}) for $\nu > (m+1)$
 \begin{eqnarray}
  \widehat{(\phi_\nu\otimes g) \psi}(k) = \int_{X} G(Y,\vx) \Theta_\nu(x_1,Y) g(x_2,\ldots,x_n)
   e^{-\I (\vx+\I\vY,k)}  d\vx  \\ \nonumber
   +(\nu +1) \int_{X} d\vx \int_{0<t<1} dt\; G(tY,\vx) e^{-\I (\vx+\I t \vY,k)} \\ \nonumber
   \cdot (\partial^\nu \phi_\nu)(x_1)
   g(x_2,\ldots,x_n) \frac{(\I Y)^\nu}{\nu!} t^\nu.
 \end{eqnarray}
 Here we used the notations $\vx=(x_1,\ldots,x_n)$ and $\vY=(Y,0,\ldots,0)$.
 With $C_2=2 e^{C_1 Y}$ we have $\vert \Theta_\nu(x,Y) \vert \leq C_2^{\nu+1}$
 and because of the bound (\ref{polybound}) we get
 \begin{equation}
  \Vert \widehat{(\phi_\nu\otimes g) \psi}(k) \Vert
  \leq C_3^{\nu+1} (e^{(Y,k)}+{(\nu-m-1)!}(-\vY,k)^{m-\nu}), \quad (\vY,k) < 0.
 \end{equation}
 We define $\psi_\nu:=(\phi_{m+\nu} \otimes g) \psi$ and if
 $(\vY,k) < -c \vert k \vert$ for a fixed $c$, we obtain for some $C_4$
 \begin{equation}
  \Vert \hat \psi_\nu (k) \Vert \leq C_4^{\nu+1} {\nu !} \vert k \vert^{-\nu},
 \end{equation}
 since $e^{-c \vert k \vert} \leq {\nu !}(c \vert k \vert)^{-\nu}$.
 If we chose $\phi_\nu$ bounded in $\co$, then $\psi_\nu$ is bounded
 in $\DiE(X,\Hilbert)$. We have shown that
 $$
  \WF_A(\psi) \subset X \times \{ k, (\vY,k) \geq 0 \}.
 $$
\end{proof}

An immediate corollary of this theorem is

\begin{cor} \label{dcone}
 Let $X \subset \rz^n$ be an open subset.
 Suppose that $V$ is an open cone in $\rz^n \backslash \{0\}$
 and $Z$ is an open neighbourhood of $0$ in $\rz^n$.
 Denote by $V^\circ$ the dual cone.\footnote{{\rm The dual cone
     $V^\circ$ of an open cone $V$ is defined as the set $V^\circ =
     \{\xi \in \rz^n: \langle v,\xi \rangle \ge 0\ \forall \, v \in
     V\}$, where $\langle\,.\,,\,.\,\rangle$ denotes the Euclidean
     scalar product on $\rz^n$.}}
 If $\psi \in \Di(X,\Hilbert)$ is the
 boundary value in the sense of distributions
 of a function $G$ which is analytic in $X \times (V
 \cap Z)$, then $\textrm{WF}_A(\psi) \subset X \times V^\circ$.
\end{cor}

\section{Quantum field theory on curved spacetimes}

By a spacetime we mean in the following a connected smooth
manifold of dimension $n \geq 2$ which is second countable
and Hausdorff and which is endowed with a Lorentzian metric $g$
such that $M$ is both oriented and time-oriented.
A spacelike hypersurface $\mathcal{C}$ in a spacetime $M$ is called
Cauchy surface 
if  each inextendible causal
curve intersects with $\mathcal{C}$ exactly once. In case there exists a
Cauchy surface the spacetime $M$ is said to be globally hyperbolic
(see e.g.\ \cite{Wald:1984rg,HawkingEllis} for further discussion).

\subsection{Scalar fields on curved spacetimes}

The Borchers-Uhlmann algebra $\mathcal{B}$ of a manifold $M$ is
defined to be the topological tensor algebra
\begin{gather}
 \mathcal{B}:=\cz \oplus \bigoplus_{m=1}^\infty {\bigotimes}^m \co(M)
\end{gather}
endowed with a star defined by
$(f_1 \otimes \ldots \otimes f_k)^*=\overline f_k \otimes \ldots \otimes
\overline f_1$.
A state $\omega$ over $\mathcal{B}$ determines a sequence of distributions
$\omega_m \in \Di(M^m)$, the so called $m$-point functions, by
 \begin{gather}
   \omega_m(f_1,\ldots,f_m):=\omega(f_1 \otimes \cdots \otimes f_m).
 \end{gather}
 If $\Hilbert$ is a Hilbert space and $D$ a dense subset we denote
 by $\mathcal{L}_D^+$ the set of (possibly unbounded) operators $A$ on
 $\Hilbert$ with the properties
 \begin{gather}
  \mathrm{dom}(A) = D,\; A D \subset D,\; \mathrm{dom}(A^*) \supset D,\; A^* D
  \subset D.
 \end{gather}
 The involution $A^+:=A^* \vert_D$ and the locally convex topology
 defined by the seminorms $p_{\phi,\psi}(A):=\vert \langle \phi, A
 \psi \rangle \vert,\; \phi,\psi \in D$ turn $\mathcal{L}_D^+$ into
 a locally convex topological $*$-algebra.

 Each state over the Borchers-Uhlmann algebra $\mathcal{B}$ determines, by the
 GNS construction, a Hilbert space $\Hilbert$ with a dense domain $D$
 and a $*$-representation $\pi: \mathcal{B} \to \mathcal{L}_D^+$ with cyclic
 vector $\Omega \in D$ such that $\pi(\mathcal{B}) \Omega =D$.
 If $M=\rz^4$ is the Minkowski space it is well known that Wightman fields
 can be constructed from states over the Borchers-Uhlmann algebra
 which satisfy certain requirements like translation invariance
 or the spectrum condition. The field is, in this case,
 the operator valued distribution
 $f \mapsto \Phi(f):=\pi(0 \oplus f \oplus 0 \oplus \cdots)$. 
 We will think of a quantum field
 on a curved spacetime in the same way, i.e.\ a quantum field
 can be defined  by a state over the Borchers-Uhlmann algebra of test functions
 on the underlying spacetime. A state is called quasifree if all the
 odd $m$-point functions vanish and the even $m$-point functions can be
 expressed by 
 \begin{gather}\label{tmf}
  \omega_m(f_1,\ldots,f_m)=\sum_P \prod_r
  \omega_2(f_{(r,1)},f_{(r,2)}), 
 \end{gather}
 where $P$ denotes a partition of the set $\{1,\ldots,m\}$ into subsets
 which are pairings of points labeled by $r$.

 For quantum fields on Minkowski spacetime one usually requires the
 properties of Poincar\'e covariance, spectrum condition, existence of
 an invariant vacuum vector as well as local commutativity (for
 observable fields) to hold (see e.g.\ \cite{SW,Haag:1992hx}).
 Due to the lack of an analogue of the Poincar\'e group,
 only the last requirement can straightforwardly be generalized to
 curved spacetimes. However, as will be seen in section \ref{mscs}, 
 there is a microlocal 
 version of the spectrum condition which can be stated independently
 of the coordinate system and hence can be applied to quantum fields
 on curved spacetimes.

 For a field $\Phi(\cdot)$ defined by a state $\omega$ over the
 Borchers-Uhlmann algebra we can associate a net of von Neumann
 algebras $\algnet$ in the following way.
 Let as above be $D$ the dense domain $\pi(\mathcal{B}) \Omega$
 arising by the GNS construction from $\omega$.
 For a subset $S \subset \mathcal{L}_D^+$ the weak commutant
 $S'_w$ of $S$ is defined to be the set of bounded operators $A$ on $\Hilbert$
 such that
 \begin{gather}
  \langle B^* \phi, A \psi \rangle = \langle A^* \phi, B \psi \rangle, \quad
  \forall\ B \in S,\; \phi,\psi \in D.
 \end{gather}
 A net of von Neumann algebras $\algnet$
 is then defined by
 \begin{gather}
  \alg(\openset):= (\{ \Phi(f); \; \mathrm{supp}(f) \subset \openset\}_w')'.
 \end{gather}
The requirement of local commutativity may now be formulated by
demanding that the von Neumann algebras associated with causally
separated regions
commute, i.e.,
$$ \alg(O_1) \subset \alg(O)' \quad \mbox{if} \quad  O_1 \subset
O^\perp\,. $$
This is a strong form of local commutativity which is to be seen as a
selective constraint on the Hilbert-space representation $\pi$ of the
Borchers-Uhlmann algebra $\mathcal{B}$ induced by the state $\omega$,
and hence as a  constraint on $\omega$ itself. It implies in
particular spacelike commutativity of field operators,
$[\Phi(f),\Phi(h)]=0$ whenever the supports of $f$ and $h$ are
causally separated.
 
\subsection{The Klein-Gordon field on curved spacetimes}

 Since the construction of free fields relies heavily on the presence
 of a Cauchy surface (time-zero formalism) we restrict our considerations
 of quantum fields on curved spacetimes to the globally hyperbolic case.
 The evolution of the free scalar field of mass $m$ and with coupling
 $\kappa$ on a globally hyperbolic
 spacetime is described by the Klein-Gordon equation
 \begin{equation}
   P \phi := (\square_g +m^2 + \kappa R) \phi =0, \quad \phi \in C^\infty(M).
 \end{equation}
 Here $\square_g$ is the Laplace operator with respect
 to the metric and $R$ is the scalar curvature of $M$.
 The operator $P$ is a differential operator of second order
 acting on the smooth functions on $M$.
 It has unique advanced and retarded fundamental solutions (see
 \cite{Leray,Dimock:1980hf}) 
 $E^\pm: C^\infty_{0}(M) \to C^\infty(M)$
 satisfying
 $$ P E^\pm = E^\pm P = \mathrm{id} \quad \textrm{on}
 \quad C^\infty_0(M)\;,$$
 $$ \textrm{supp}(E^\pm f) \subset J^\pm(\textrm{supp}(f))\;,$$
 where $J^\pm(\openset)$ denotes the causal future/past of a set $\openset$,
 i.e. the set of points which can be reached by future/past directed causal
 curves emanating from $\openset$. The difference $E:=E^+ - E^-$
 is the so called commutator function. It maps $\co(M)$ onto the space
 of smooth solutions to the Klein-Gordon equation which have compactly
 supported restriction to all Cauchy surfaces. The map
 $E: \co(M) \to C^\infty(M)$ is continuous and hence has a
 distributional kernel in $\Di(M \times M)$ which we also denote by $E$,
 i.e. $E(f,h)=\int_M f (Eh)$ where integration is taken with
 respect to the Lorentzian metric-volume form. The field algebra $\feld$ 
 of the Klein-Gordon
 field is defined to be the unital $*$-algebra generated by the symbols
 $\phi(f), \; f \in \co(M)$ and the relations
 \begin{enumerate}
  \item $f \to \phi(f)$ is complex linear,
  \item $\phi(f)^*=\phi(\overline f)$,
  \item $\phi(P f)=0$,
  \item $[\phi(f_1),\phi(f_2)]=\I E(f_1,f_2)$.
\end{enumerate}
Clearly, each state $\omega$ over $\feld$ with the further property that
the $\omega(\phi(\cdot) \cdots \phi(\cdot))$ are distributions
defines a state $\tilde \omega$ over the Borchers-Uhlmann algebra by
\begin{gather}
  \tilde \omega(f_1 \otimes \ldots \otimes f_m):=
  \omega(\phi(f_1) \cdots \phi(f_m)).
\end{gather}
The corresponding quantum field $\Phi: \co(M) \to
\mathcal{L}_D^+$ satisfies the Klein-Gordon equation
\begin{gather}
 \Phi(P\; \cdot) = 0,
\end{gather}
and the canonical commutation relations
\begin{gather}\label{ccr}
 [\Phi(f_1),\Phi(f_2)]=\I E(f_1,f_2) \mathbb{I}\vert_{D}.
\end{gather}
Since the commutator function $E$ vanishes for spacelike
separation of the arguments the field satisfies the requirement of
local commutativity, i.e. $[\Phi(f),\Phi(h)]=0$ if the supports of $f$
and $h$ are spacelike separated. For many states, among them
the quasifree states, 
also the stronger requirement of local commutativity at the level of
the net of von Neumann algebras $\algnet$ in their GNS-representations
described above is fulfilled.
In the following we call states over the Borchers-Uhlmann algebra
arising in this way states for the Klein-Gordon field.

\section{The microlocal spectrum condition} \label{mscs}

In the investigation of the Klein-Gordon field a crucial role
is played by the so called Hadamard states 
(see e.g. \cite{Fulling:1989nb,Kay:1991mu,Wald}).
They are thought of as the appropriate counterpart of the vacuum in Minkowski
space and are characterized by the short distance behaviour of their
$2$-point function. 
The investigation of such states is partially motivated by the result of Wald
(\cite{Wald:1977up}) that
the expectation value of the energy momentum tensor $T_{\mu\nu}$ 
with respect to a
Hadamard state can be made sense of in a satisfactory way. 
This is a very important feature of
Hadamard states, since it is this expectation value that appears in the
Einstein equations in the semi-classical theory of gravity coupled to the
Klein-Gordon field.
We will not give the original definition of Hadamard states here, since such
states can as well be characterized by the wavefront set of their
$2$-point function. 
This was shown by Radzikowski (\cite{Radzikowski:1992eg,Radzikowski:1996pa})
and relies heavily on the work of Duistermaat and H\"ormander
(\cite{Hormander:1971,Hormander:1972}) on Fourier integral operators.
We first would like to note that the wavefront set of the commutator
distribution $E$ for the wave operator $P$ on a globally hyperbolic
spacetime with metric $g$
is given by
\begin{gather}
 \textrm{WF}(E)=\{ \left ((x_1,-k_1),(x_2,k_2)\right) ; (x_1,k_1) \sim (x_2,k_2)\\\nonumber
 \textrm{ and } g^{\mu\nu}(k_1)_{\mu}(k_1)_\nu=
 g^{\mu\nu}(k_2)_{\mu}(k_2){\nu}=0 \}, 
\end{gather}
where $(x_1,k_1) \sim (x_2,k_2)$ means that there is a lightlike geodesic $\gamma$
connecting $x_1$ and $x_2$ such that $k_1$ is coparallel to the tangent
vector of the curve at $x_1$ and $k_2$ is the parallel transport of $k_1$
from $x_1$ to $x_2$.
Radzikowski's result is that a quasifree state $\omega$ for the
Klein-Gordon field is a 
Hadamard state if and only if the wavefront set of its $2$-point function $\omega_2$
is given by
\begin{gather}\label{Had}
  \textrm{WF}(\omega_2)=\{ \left((x_1,-k_1),(x_2,k_2)\right) \in
  \WF(E) ; (k_2)_0 > 0 \}\,,
 \end{gather}
where $(k_2)_0 > 0$ is shorthand for ``$k_2$ is future-pointing''.

The microlocal characterization of Hadamard states has meanwhile led to a
rich theory. In fact, it turned out that quasifree Hadamard states
allow for a construction of Wick polynomials
of field operators
(\cite{Brunetti:1996rf,HollandsW:2001})
and a perturbative construction of interacting fields
on curved spacetimes (\cite{Brunetti:1999jn,Hollands:2001fb}).
It was shown in \cite{Verch:1994eg} that such states are
locally quasiequivalent and therefore (at least locally) distinguish a single
folium of states. A passive state for a free quantum field theory
on a stationary spacetime is always a Hadamard state (see
\cite{Fulling:1981cf} and \cite{Verch:vs00} for
the statement in its full generality). The Hadamard condition can also be
formulated almost without change
for arbitrary free quantum fields. Radzikowski's result is known
to hold also in these cases (\cite{Sahlmann:2000ij}). Adiabatic vacuum
states satisfy a similar condition with
the wavefront set replaced by an appropriate Sobolev-wavefront set as shown
in \cite{Junker:2001gx}.

Motivated by the observations just mentioned, a microlocal spectrum
condition (\mbox{\msc})  that applies to general quantum fields
on curved spacetimes was 
introduced by Brunetti, Fredenhagen and K\"ohler 
\cite{Brunetti:1996rf}; we shall now summarize its definition.
  We denote by $\mathcal{G}_k$ the set of all finite
graphs with vertices $\{ 1,\ldots,k\}$
such that for every element $G \in \mathcal{G}_k$ all edges occur in both
admissible directions. We write $s(e)$ and $r(e)$ for the source and the
target of an edge respectively. Following \cite{Brunetti:1996rf} we define an
immersion of a graph $G \in \mathcal{G}_k$ into a spacetime $M$ an assignment
of the vertices $\nu$ of $G$ to points $x(\nu)$ in $M$, and of edges $e$
of $G$ to piecewise smooth curves $\gamma(e)$ in $M$ with source
$s(\gamma(e))=x(s(e))$ and range $r(\gamma(e))=x(r(e))$, together
with a covariantly constant causal covector field $k_e$ on $\gamma$
such that
\begin{enumerate}
 \item If $e^{-1}$ denotes the edge with opposite direction as $e$, then
   the corresponding curve $\gamma(e^{-1})$ is the inverse of $\gamma(e)$.
 \item For every edge $e$ the covector field $k_e$ is directed towards the
   future whenever $s(e) < r(e)$.
 \item $k_{e^{-1}}=-k_e$.
\end{enumerate}

\begin{definition}[\mbox{\msc},\;\cite{Brunetti:1996rf}] \label{MSC}
 A state $\omega$ over the Borchers-Uhlmann algebra $\mathcal{B}$
 is said to satisfy the {\sl microlocal spectrum condition}
 iff its $m$-point functions $\omega_m \in \Di(M^m)$ satisfy
 \begin{gather*}
  \WF(\omega_m) \subset \lbrace
   (x_1,k_1; \ldots; x_m,k_m) \in T^*M^m \backslash 0 ; \;
   \exists G \in \mathcal{G}_m\\ \textrm{ and an immersion } (x,\gamma,k)
   \textrm{ of } G \textrm{ in }, \textrm{ such that } \\
   x_i = x(i) \quad \forall i = 1,\ldots,m \textrm{ and }\\
   k_i=-\sum_{e,\;s(e)=i} k_e(x_i)
  \} = \Gamma_m
 \end{gather*}
\end{definition}

We note here that quasifree Hadamard states of the Klein-Gordon field
satisfy the microlocal spectrum condition.
For a motivation of this definition and further properties of states satisfying the
\mbox{\msc} we refer the reader to \cite{Brunetti:1996rf}. For later purposes we
will need the following property of the sets $\Gamma_m$ which is Lemma 4.2 in \cite{Brunetti:1996rf}.
\begin{pro}\label{sua}
 The sets $\Gamma_m$ are stable under addition, i.e. $\Gamma_m + \Gamma_m
 \subset \Gamma_m$. Moreover, if $(x,k) \in \Gamma_m$ then $(x,-k) \notin
 \Gamma_m$.
\end{pro}

\section{The analytic microlocal spectrum condition and the Reeh-Schlieder property}

In the following we restrict our consideration to the case when $M$ is a real
analytic spacetime, i.e.\ $M$ is real analytic as a manifold and the metric $g$
is analytic. Passing from the smooth to the analytic category it seems
reasonable to require that the state satisfies a microlocal spectrum
condition with the wavefront set $\WF$ replaced by the analytic
wavefront set $\WF_A$. (See \cite{HollandsW:2001} for a related concept.)
\begin{definition}[\amsc] \label{AMSC}
 A state $\omega$ over the Borchers-Uhlmann algebra is said to satisfy the
 {\sl Analytic Microlocal Spectrum Condition} (\amsc) iff its $m$-point
 functions satisfy
 \begin{gather*}
  \WF_A(\omega_m) \subset \Gamma_m,
 \end{gather*}
 where the notations of Def. \ref{MSC} are used.
\end{definition}

For Wightman fields in Minkowski spacetime the spectrum condition is
equivalent to the requirement that the $m$-point functions are boundary
values of functions which are analytic in the tube
\begin{gather}
  T_m:=\{(z_1,\ldots,z_m);\; \Im(z_{j+1}-z_j) \in V_+,
  j=1,\ldots,m-1\},
\end{gather}
where $V_+$ is the forward lightcone.
\begin{theorem}
 Suppose that $M$ is the $n$-dimensional Minkowski spacetime ($n \geq 2$)
 and let $\omega$ be a state over the Borchers-Uhlmann algebra such that
 its $m$-point functions are boundary values in the sense of distributions of functions
 that are analytic in $T_m \cap Z$, where $Z$ is a complex
 neighbourhood of $(\rz^n)^m \subset (\cz^n)^m$. Then $\omega$ satisfies
 \mbox{\amsc}.
\end{theorem}
\begin{proof}
 Clearly, $T_m$ is of the form $T_m = (\rz^n)^m + \I \mathcal{C}$,
 where the cone $\mathcal{C}$ is defined by
 $\mathcal{C}=\{(k_1,\ldots,k_m);\; k_{j+1}-k_j \in V_+\}$. 
 The dual cone $\mathcal{C}^\circ$ can easily be calculated and the result is
 \begin{gather}
   \mathcal{C}^\circ=\left\{(k_1,\ldots,k_m);\; k_m,k_{m-1}+k_m,\ldots,\sum_{j=2}^{m}
   k_j \in \overline V_+, \sum_{j=1}^m k_j=0 \right\}.
 \end{gather}
 Hence, by Corollary
 \ref{dcone}, the set $\WF_A(\omega_m)$ is contained in
 $(\rz^n)^m \times \mathcal{C}^\circ$.
 The set
 $$
  \left\{(x_1,k_1;\ldots;x_m,k_m);\; k_m,k_{m-1}+k_m,\ldots,\sum_{j=2}^{m}
   k_j \in \overline V_+, \sum_{j=1}^m k_j=0 \right\}
 $$
 is contained in $\Gamma_m$ (see e.g. the proof of Theorem 4.6 in
 \cite{Brunetti:1996rf}) which concludes the proof.
\end{proof}
This theorem applies to the vacuum state of Wightman fields in Minkowski
spacetime 
and, provided that the invariant domain of all field operators includes e.g.\
the $C^\infty$-vectors for the energy, it applies also to vector states
which are analytic in the energy (in vacuum representation, see e.g.\
Chp.\ 12 in \cite{BW}); moreover,
it applies also
to states which satisfy the relativistic KMS condition proposed by Buchholz
and Bros (\cite{Bros:1994ua,Bros:1996mw}).
Quasifree states for the Klein-Gordon field on the de Sitter spacetime
that satisfy the weak spectral condition
(\cite{Bros:1996js,Bros:1998ik,Bros:2001yw}) 
can also be shown to satisfy \mbox{\amsc}.

Let as before $(\pi,\Omega,D,\Hilbert)$ be the GNS-representation of the state
$\omega$ and denote by $\Phi$ the corresponding quantum field.
We will show now that a state that satisfies \mbox{\amsc} has the
Reeh-Schlieder 
property, i.e. the set $\{ \Phi(f_1) \cdots \Phi(f_n) \Omega; \; \supp(f_i)
\subset \openset, m \in \nz \}$ is total in $\Hilbert$ for each
non-void open set $\openset \subset M$. 

The main technical tool for proving this is a microlocal version of the
{\sl edge of the wedge} theorem (Thm.\ 8.5.6' in \cite{HormBook:1990}).
\begin{pro} \label{eotw}
 Let $M$ be a real analytic connected manifold and $u \in \Di(M)$ a distribution with
 the property that $\WF_A(u) \cap -\WF_A(u) = \emptyset$. Then the
 following conclusion holds for each non-void open subset $\openset \in M$:
 \begin{gather*}
  u\vert_{\openset} = 0 \implies u=0.
 \end{gather*}
\end{pro}
\begin{proof}
 For a closed subset $X \subset M$ the exterior normal set $N_e(X) \subset
 T^*M \backslash 0$ is defined to be the set of all $(x,k)$ such that $x \in
 X$ and there is a real valued function $f \in C^2(M)$ with $df(x)=k \not= 0$
 and $f(y) \leq f(x)$ for all $y \in X$. The normal set $N(X)$ is the union
 $N_e(X) \cup -N_e(X)$.
 Theorem 8.5.6' in \cite{HormBook:1990}
 states that $N(\supp(u)) \subset \WF_A(u)$ and since
 $N(\supp(u)) = -N(\supp(u))$ the assumption implies that $N(\supp(u))
 = \emptyset$. Consequently, $N_e(\supp(u)) = \emptyset$. Prop. 8.5.8
 in \cite{HormBook:1990} states that the projection of $N_e(X)$
 in $M$ is dense in $\partial X$. Therefore, $\partial(\supp(u)) =
 \emptyset$. Since $M$ is connected this implies that either
 $\supp(u) = \emptyset$ or $\supp(u) = M$. The latter is excluded
 by $u \vert_{\openset} = 0$.
\end{proof}

Our main result is stated in the following theorem.

\begin{theorem}\label{RST}
 Let $\omega$ be a state over the Borchers-Uhlmann algebra on a real analytic
 spacetime and suppose furthermore
 that $\omega$ satisfies \amsc. Denote by $(\pi,\Omega,D,\Hilbert)$ its
 GNS-representation and by $\Phi$ the associated quantum field.
 Then the set
 $$\{ \Phi(f_1) \cdots \Phi(f_m) \Omega; \; \supp(f_i) \subset \openset, m
 \in \nz \}$$
is total in $\Hilbert$ for each non-void open set $\openset \subset M$.
\end{theorem}
\begin{proof}
 We define a Hilbert space valued distribution $\psi_m \in \Di(M^m,\Hilbert)$
 by
 \begin{gather}
  \psi_m(f_1,\ldots,f_m):=\Phi(f_1) \cdots \Phi(f_m) \Omega.
 \end{gather}
 Note that due to Prop. \ref{m2} a point $(x,k) \in T^*M^m \backslash 0$
 is in $\WF_A(\psi)$ if and only if $(x,-k;x,k) \in \WF_A(w_{2m})$,
 where the distribution $w_{2m} \in \Di(M^{2m})$ is defined by
 \begin{gather}
  w_{2m}(f_1,\ldots,f_m,g_1,\ldots,g_m):=\omega_{2m}(f_m,\ldots,f_1,g_1,\ldots,g_m).
 \end{gather}
 Prop. \ref{sua} implies that $WF_A(w_{2m}) \cap -WF_A(w_{2m})=\emptyset$
 and therefore $WF_A(\psi_m) \cap -WF_A(\psi_m)=\emptyset$.

 Suppose now that $\phi \in \Hilbert$ is orthogonal to the set
 \begin{gather}
  \{ \Phi(f_1) \cdots \Phi(f_m) \Omega; \; \supp(f_i) \subset \openset, m \in
  \nz\}.
 \end{gather}
 Then the distributions $v_m(\cdot)=\langle \phi,\psi_m(\cdot) \rangle \in
 \Di(M^m)$ vanish on $O^m$ and satisfy $WF_A(v_m) \cap -WF_A(v_m)=\emptyset$.
 By Prop. \ref{eotw} we conclude that $v_m=0$ for all $m$. Therefore,
 $\phi$ is even orthogonal to the set
 \begin{gather}
  \{ \Phi(f_1) \cdots \Phi(f_m) \Omega; \; f_i \in \co(M), m \in \nz\},
 \end{gather}
 which is total in $\Hilbert$. 
We conclude that $\phi=0$ which proves the theorem.
\end{proof}

An immediate corollary is the Reeh-Schlieder property of the associated
net of local algebras.
\begin{cor}
 Let the assumptions of theorem \ref{RST} be fulfilled and denote by
 $\algnet$ the associated local net of von Neumann algebras.
 Then $\alg(\openset) \Omega$ is dense in $\Hilbert$ for each non-void
 open set $\openset$.
\end{cor}

\begin{rem}
 Clearly, the conclusion of theorem \ref{RST} also holds if we impose
 the weaker condition $WF_A(\omega_{2m}) \cap -WF_A(\omega_{2m})=\emptyset$
 on the state instead of the analytic microlocal spectrum condition.
 Our result is therefore insensitive to the precise form of analytic microlocal
 spectrum condition as long as an analogue of Prop. \ref{sua} holds.
\end{rem}

\begin{rem}
 The same method works for fields with values in an analytic vector bundle.
 Note that local commutativity is not an assumption of Thm.\ \ref{RST}
 and therefore it applies to fermionic fields as well.
\end{rem}

\section{Quasifree states and the Klein-Gordon field}

As indicated in the introduction the microlocal spectrum
condition can be simplified for quasifree states of the Klein-Gordon field
using microlocal analysis of Hilbert space valued distributions.

\begin{pro} \label{simpl}
 Let $M$ be a globally hyperbolic spacetime and let $\omega$ be a quasifree state for the
 Klein-Gordon field on $M$.
 Let $\Phi$ and $\Omega$ be as in the previous
 section and denote by $\psi$ the Hilbert space valued distribution
 $\Phi(\cdot) \Omega$. Then the following statements are equivalent.
 \begin{itemize}
  \item[1)] $\omega$ satisfies \mbox{\msc}.
  \item[2)] $\WF(\psi) \subset \overline V_+$.
  \item[3)] $\WF(\psi) = N_+$.
  \item[4)] $\omega$ is a Hadamard state.
 \end{itemize}
 Here $\overline V_+$ denotes the set of future directed
 causal covectors $(x,k)$ and $N_+$ is the set of future directed
 non-zero null-covectors.
\end{pro}

\begin{proof}
  We first show that 1) $\Rightarrow$ 2).
  Suppose that $(x,k) \in \WF(\psi)$. Then by Prop. \ref{m1} we get
  $(x,-k;x,k) \in \WF(\omega_2)$. By \mbox{\msc} the covector $(x,k)$
  must be in $\overline V_+$.  2) $\Rightarrow$ 3) is a simple consequence
  of the fact that $\psi$ solves the Klein-Gordon equation and
  Prop. \ref{diffwf}. The implication 4) $\Rightarrow$ 1) is Prop. 4.3
  in \cite{Brunetti:1996rf} and it remains to show 3) $\Rightarrow$ 4).
  By Prop. \ref{m1} we conclude that if $(x_1,k_1;x_2,k_2) \in \WF(\omega_2)$
  then $k_2 \in N_+$ and $k_1 \in N_-$. Therefore,
  $\WF(\tilde \omega_2) \cap \WF(\omega_2) = \emptyset$ with
  $\tilde\omega_2(f_1,f_2):=\omega_2(f_2,f_1)$. Note
  that $\tilde \omega_2 - \omega_2$ is proportional to the commutator function $E$
  and consequently $\WF(\omega_2) \cup \WF(\tilde \omega_2)= \WF(E)$.
  This implies Eq. (\ref{Had}) and thus concludes the proof.
\end{proof}

A completely analogous statement holds in the analytic category.
\begin{pro} \label{asimpl}
 Let $M$ be an globally hyperbolic analytic spacetime and let $\omega$ be a quasifree state for the
 Klein-Gordon field on $M$. With the same notation as in theorem \ref{simpl}
 the following statements are equivalent.
 \begin{itemize}
  \item[1)] $\omega$ satisfies \mbox{\amsc}.
  \item[2)] $\WF_A(\psi) \subset \overline V_+$.
  \item[3)] $\WF_A(\psi) = N_+$.
  \item[4)] $\omega$ is an analytic Hadamard state, meaning that
    (\ref{Had}) is satisfied with $\WF$ replaced by $\WF_A$.
  \end{itemize}
 \end{pro}
\begin{proof}
 Taking into account Prop. \ref{m2} and Prop. \ref{adiffwf} the proof is identical
 to that of Prop. \ref{simpl} with $\WF$ replaced by $\WF_A$.
\end{proof}

A spacetime $M$ with metric tensor $g$ is called stationary if there
exists a $1$-parameter group $h_t$ of isometries of $M$ with timelike
orbits whose Killing vector fields are (by convention) future pointing.
This $1$-parameter group can be understood as a group of time-translations
and it is therefore interesting to investigate passive states with respect
to this group action as states which are physically reasonable replacements
for the vacuum. For example the Schwarzschild spacetime is stationary and
the Hartle-Hawking state for the Klein-Gordon field is a KMS state with
respect to the group of time translations. Other examples are the Rindler
wedge and wedge-like regions in the de Sitter space. We investigate in the following
ground and KMS states for the Klein-Gordon field on a stationary spacetime.
Let us first fix some notation.
The push-forward $h_t{}_*$ defined by $(h_t{}_* f)(x):=f(h_{-t}x)$
acts on the space $\co(M)$ and this action lifts uniquely to an action
$\alpha_t$ 
on the Borchers-Uhlmann algebra $\mathcal{B}$ by $*$-automorphisms.
A state $\omega$ over $\mathcal{B}$ is called ground state
if the function $t \to \omega(A \alpha_t B)$ is bounded for all $A,B \in
\mathcal{B}$ and
\begin{gather}
 \int_{-\infty}^{+\infty} \hat f (t) \omega(A \alpha_t (B)) dt = 0,
\end{gather}
holds for all $f \in \co((-\infty,0))$.
A state $\omega$ is called KMS state at inverse temperature $\beta>0$ if
the function $t \to \omega(A \alpha_t B)$ is bounded for all $A,B \in
\mathcal{B}$ and
\begin{gather}
 \int_{-\infty}^{+\infty} \hat f (t) \omega(A \alpha_t (B)) dt =
 \int_{-\infty}^{+\infty} \hat f (t+\I \beta) \omega(\alpha_t(B) A) dt,
\end{gather}
for all $f \in \co(\rz)$.
Note that ground and KMS states are necessarily invariant,
i.e. $\omega(\alpha_t(\cdot))=\omega(\cdot)$.  

\begin{theorem}
 Let $M$ be a globally hyperbolic stationary spacetime
 and suppose that $\omega$ is a quasifree KMS- or ground-state for the Klein-Gordon
 field. Then $\omega$ satisfies the microlocal spectrum condition.
 If moreover $M$ is real analytic (as a spacetime) and the flow $\rz \times M \to M$ induced
 by $h_t$ is analytic, then $\omega$ satisfies the analytic
 microlocal spectrum condition.
\end{theorem}

\begin{proof}
  Let $(\pi,\Omega,\Hilbert,D)$ be the GNS representation of $\omega$. Let
  $\Phi(\cdot)$ be the associated field and define $\psi(\cdot):=\Phi(\cdot) \Omega$.
  Since $\omega$ is invariant there exists a strongly continuous one-parameter
  group $U(t)=e^{\I t \mathbf{H}}$ on $\Hilbert$, such that
  $U(t) \psi(\cdot) = \psi(h_t \cdot)$ and $U(t) \Omega= \Omega$.
  If $\omega$ is a $\beta$-KMS state it follows from the KMS condition that
  the vectors $\psi(f)$ are in the domain
  of $e^{-\frac{\beta}{2} \mathbf{H}}=U(\I \beta/2)$ for all $f \in \co(M)$.
  If $\omega$ is a ground state this is even true for all $\beta>0$.
  Therefore, we may define the distribution $G \in \Di((0,\beta/2) \times M,\Hilbert)$ by
  $G(t,x)=U(\I t) \psi(x)$.
  We use a local coordinate system $(x_0,\ldots,x_{n-1})$ such that the
  vector field $\partial_{x_0}$ generates locally the flow $h_t$.
  Then $G$ satisfies the system of equations
  \begin{eqnarray}
    (\partial^2_t + \partial^2_{x_0}) G = 0 \\
    (\textrm{id} \otimes P) G = 0.
  \end{eqnarray}
  This system is elliptic and therefore $G$ is indeed a smooth function
  satisfying the conditions of Theorem \ref{boundval}.
  It follows that $WF_A(\psi)$ is a subset of the set
  $\{ (x,k) \in T^*M; k(\partial_{x_0}(x))>0 \}$ in this coordinate system.
  Since $\psi$ solves the Klein-Gordon equation $\WF(\psi)$ is confined to
  the forward light cone. If $M$ is real analytic and the
  flow $\rz \times M \to M$ induced by $h_t$ is analytic we can choose the
  local coordinate system to be analytic and therefore $\WF_A(\psi) \subset
  \overline V_+$.
\end{proof}
${}$\\[20pt]
{\bf Acknowledgements}\\[6pt]
We would like to thank Otto Liess for some helpful remarks at the
beginning of this project.
R.V.\ wishes to thank the Institut f.\ Theoretische Physik and the
``Graduiertenkolleg Quantenfeldtheorie'' in Leipzig for hospitality
during Jan.--Feb.\ 2000, and a travel grant.
A.S. would like to thank the Institut f.\ Theoretische Physik in G\"ottingen
for a stay there at the beginning of May 2001 (thanks to ``Cron und Lanz''
for the fancy cakes).
This work was partially supported by the
Deutsche Forschungsgemeinschaft within the scope of the postgraduate
scholarship programme ``Graduiertenkolleg Quantenfeldtheorie'' at the
University of Leipzig.

\end{document}